# Quantum Griffiths Singularity in a Three-Dimensional Superconductor to Anderson Critical Insulator Transition

Shichao Qi[1], Yi Liu[2,3], Ziqiao Wang[1], Fucong Chen[4], Qian Li[4], Haoran Ji[1], Rao Li[2], Yanan Li[1], Jingchao Fang[1], Haiwen Liu[5,6,‡], Fa Wang[1], Kui Jin[4,7,8,†], X. C. Xie[1,6,9], Jian Wang[1,9,10,*]

[1]*International Center for Quantum Materials, School of Physics, Peking University, Beijing 100871, China*
[2] *Department of Physics and Beijing Key Laboratory of Opto-electronic Functional Materials & Micro-nano Devices, Renmin University of China, Beijing 100872, China*
[3]*Key Laboratory of Quantum State Construction and Manipulation (Ministry of Education), Renmin University of China, Beijing 100872, China*
[4]*Beijing National Laboratory for Condensed Matter Physics, Institute of Physics, Chinese Academy of Sciences, Beijing 100190, China*
[5]*Center for Advanced Quantum Studies, Department of Physics, Beijing Normal University, Beijing 100875, China*
[6]*Interdisciplinary Center for Theoretical Physics and Information Sciences, Fudan University, Shanghai 200433, China*
[7]*School of Physical Sciences, University of Chinese Academy of Sciences, Beijing 100049, China*
[8]*Songshan Lake Materials Laboratory, Dongguan, Guangdong 523808, China*
[9]*Hefei National Laboratory, Hefei 230088, China*
[10]*Collaborative Innovation Center of Quantum Matter, Beijing 100871, China*

Disorder is ubiquitous in real materials and can have dramatic effects on quantum phase transitions. Originating from the disorder enhanced quantum fluctuation, quantum Griffiths singularity (QGS) has been revealed as a universal phenomenon in quantum criticality of low-dimensional superconductors. However, due to the weak fluctuation effect, QGS is very challenging to detect experimentally in three-dimensional (3D) superconducting systems. Here we report the discovery of QGS associated with the quantum phase transition from 3D superconductor to Anderson critical insulator in a spinel oxide $MgTi_2O_4$ (MTO). Under both perpendicular and parallel magnetic field, the dynamical critical exponent diverges when approaching the quantum critical point, demonstrating the existence of 3D QGS. Among 3D superconductors, MTO shows a relatively strong fluctuation effect featured as a wide superconducting transition region. The enhanced fluctuation, which may arise from the mobility edge of Anderson localization, finally leads to the occurrence of 3D quantum phase transition and QGS. Our findings offer a new perspective to understand quantum phase transitions in strongly disordered 3D systems.

Driven by quantum fluctuations, the quantum phase transition (QPT) represents zero temperature phase transition between different quantum ground states [1,2]. The QPT has been observed in different physical systems, such as superconductors [3], quantum anomalous Hall systems [4], heavy-fermion materials [5], and ultracold atoms [6]. Understanding the dual effect of disorder and quantum fluctuation on QPTs is among the central topics in condensed matter physics. The electronic wave functions in solids are usually described as extended Bloch waves due to the periodicity of the crystal lattice. Anderson pointed out that in the strong disorder regime, the wave function may become localized and the system changes into an insulating state [7]. The transition between the extended and localized states occurs at the mobility edge [8]. Near the mobility edge, the fluctuation effect is significantly enhanced [9-11], offering new opportunity to investigate the influence of strong disorder on the critical behavior of QPT.

Benefiting from the strong fluctuation effect, low-dimensional superconducting systems are promising platforms to investigate the superconductor to insulator or metal transitions (SIT/SMT) [3,12-14] as a paradigm of QPT. Quantum Griffiths singularity (QGS) of SMT has been discovered in Ga films [15,16] and subsequently detected in various low-dimensional superconductors [17-23]. Distinct from conventional SIT/SMT [3,13], the prominent feature of QGS is the divergence of dynamic critical exponent $z$ when approaching the infinite randomness quantum critical point [24,25]. The divergent $z$ indicates ultraslow dynamics of superconducting rare regions [24,25], reflecting the profound influence of quenched disorder on SMT in low-dimensional superconductors. Nevertheless, the experimental detection of QGS in 3D superconductors is very challenging due to the relatively weak fluctuation effect. Theoretically, the superconductor to fermionic insulator or metal transition (fermionic SIT/SMT) in the clean limit can only occur in 1D or 2D systems [13]. It was proposed the strong disorder can induce strong fluctuation effect near the mobility edge of 3D Anderson localization and thus stabilize a fermionic SIT in 3D systems [11], but experimental evidence is scarce. Given the strong disorder and fluctuation effect near the mobility edge, it is promising to study the quantum criticality of 3D SIT in the Anderson critical regime (also called the Anderson critical insulator).

In this Letter, we report the discovery of QGS in a 3D superconductor to Anderson critical insulator transition in spinel oxide MgTi$_2$O$_4$ (MTO). The crystalline MTO films were grown on MgAl$_2$O$_4$(00$l$) (MAO) substrates via pulsed laser deposition [26] (see Supplemental Material for details [27]). The thicknesses of sample s1 and s2 are 377 and 474 nm, respectively. The inset of Fig. 1(a) presents the schematic crystal structure of MTO (see Fig. S1(a) for another oblique view of the crystal structure). The x-ray diffraction results demonstrate that the ab-plane of MTO is parallel to that of the MAO substrate (Fig. S1(b)). The high-resolution scanning transmission electron microscopy image illustrates the spinel structure of MTO with lattice constants of $a$ (or $b$) = 0.86 nm and $c$ = 0.89 nm (Fig. S1(d)). For transport measurements, the standard four-electrode method (inset of Fig. 1(b)) is used to investigate the superconducting properties of MTO. With decreasing temperature, the temperature dependent resistivity $\rho(T)$ for sample s1 shows an insulating behavior followed by a superconducting transition at low temperatures (Fig. 1(a)). The resistivity starts to drop at $T_c^{\text{onset}} = 5.60$ K and reaches zero within the measurement resolution at $T_c^{\text{zero}} = 2.88$ K (Fig. 1(b)). Here $T_c^{\text{onset}}$ is defined as the temperature where the resistivity deviates from the linear extrapolations of the normal state. Figures 1(c) and 1(d) present resistivity under perpendicular (perpendicular to the $ab$ plane) and parallel (parallel to the $ab$ plane) magnetic fields at temperatures from 0.1 to 4.0 K, respectively. The superconductivity can be largely suppressed under 10 T at low temperatures, exhibiting almost isotropic feature of the upper critical field. As shown in Fig. 1(e), the perpendicular and parallel upper critical fields, defined as the fields corresponding to 50% of the normal state resistivity ($\rho_n$), show subtle difference. This is further confirmed by magnetoresistivity measurements at different orientations from perpendicular (0°) to parallel (90°) directions at 3 K on sample s2 (Fig. 1(f)). Experimentally, the ratio of the perpendicular upper magnetic field to the parallel one ($B_{c2}^{\perp}/B_{c2}^{//}$) is a common way to reveal the dimensionality of superconductivity [66-69]. For 2D superconductors, $B_{c2}^{//}$ is much larger than $B_{c2}^{\perp}$. However, $B_{c2}^{//}$ of MTO is even slightly smaller than $B_{c2}^{\perp}$, implying the characteristics of 3D superconductivity. The main evidence of 3D superconductivity in MTO is provided by comparing the film thickness and superconducting coherence length. According to the Ginzburg-Landau (GL) theory [70], the GL coherence length $\xi_{\text{GL}}$ of Cooper pairs can be derived by the formula $B_{c2} = \frac{\Phi_0}{2\pi \xi_{\text{GL}}^2}(1 - \frac{T}{T_c})$, where $\Phi_0 = h/2e$ is the flux quantum. The fitting of the temperature dependent $B_{c2}^{\perp}$ ($B_{c2}^{//}$) near $T_c$ yields $\xi_{\text{GL}} = 2.85$ nm (3.60 nm), 2 orders of magnitude smaller than the film thickness $d$ of 377 nm (s1), providing direct evidence of 3D superconductivity. Note that for 2D superconductors, the coherence length is usually larger than (or at least comparable to) the thickness (Table S1). Therefore, the large ratio $d/\xi_{GL} > 100$, in combination with the nearly isotropic critical fields, reveals the 3D superconducting nature of MTO. Furthermore, the 3D spinel structure of MTO can rule out the possibility of a layer-stacked structure, which further reinforces the conclusion of 3D superconductivity. Moreover, both the $B_{c2}^{\perp}$ and $B_{c2}^{//}$ can be well fitted by the Werthamer-Helfand-Hohenberg (WHH) formula [71] (Fig. 1(e)), yielding the impurity scattering strength $\alpha = \frac{3}{2k_F l} = 5.29$ (3.41) and the Ioffe-Regel parameter $k_F l = 0.28$ (0.44) for perpendicular (parallel) fields [27]. Here, $k_F$ is the Fermi wave vector and $l$ is the mean free path. The small value of $k_F l$ suggests the MTO is a strongly disordered system and the normal state is near the mobility edge of Anderson localization [72,73]. The strong disorder may originate from the lattice distortions and atom vacancies of the MTO film [27].

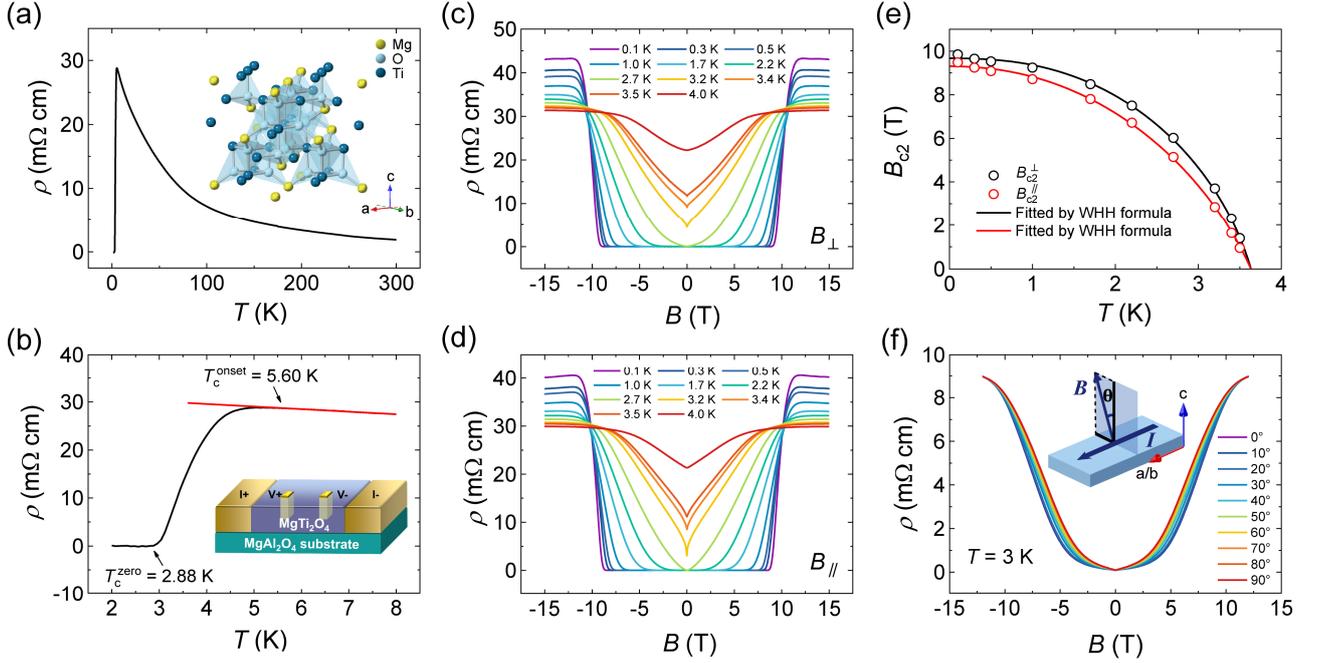

FIG. 1. 3D superconductivity of MTO. (a) Temperature dependence of resistivity $\rho$ at zero field. The inset shows the spinel crystal structure of MTO. The yellow, cyan and dark blue spheres represent the magnesium, oxygen and titanium atoms, respectively. (b) The enlarged view of $\rho(T)$ curve at zero field with $T_c^{onset} = 5.60$ and $T_c^{zero} = 2.88$ K. Inset: Schematic for standard four-electrode transport measurements. (c) (d) Perpendicular (c) and parallel (d) magnetoresistivity isotherms from 0.1 to 4.0 K. (e) Temperature dependence of perpendicular (black spheres) and parallel (red spheres) critical fields. The solid lines are fittings by the WHH formula. (f) The magnetoresistivity measured at different orientations at $T = 3$ K. Inset: Schematic of sample geometry. The excitation current is applied along $a/b$ axis. The magnetic field is always perpendicular to the current during the rotation. $\theta$ represents the angle between the magnetic field and $c$ axis. The data in panels (a)–(e) is from sample s1 and that in panel (f) is from s2.

Under perpendicular field, the superconductivity of MTO is gradually suppressed and the system undergoes a superconductor to insulator transition (Fig. 2(a)). The temperature dependence of resistance in the insulating state of MTO significantly deviates from the theoretical prediction of the bosonic insulating state (Fig. S3(a)), revealing the dominant contribution of fermionic quasiparticles [27]. Consistently, the large impurity scattering strength $\alpha$ derived by WHH formula (Fig. 1(e)) reveals that the superconductivity is destroyed by the spin pair-breaking effect [71], corresponding to the fermionic SIT. Moreover, when the superconductivity is fully eliminated by the perpendicular field of 15 T, the $\rho(T)$ curve at low temperatures shows the fermionic insulating behavior, since it is quite similar to the $\rho(T)$ behavior of fermionic normal state above $T_c^{onset}(0\ T)$ (Fig. S3(b)). To investigate the quantum critical behavior of SIT in MTO, we systematically measured the magnetoresistivity isotherms at ultralow temperatures from 50 to 600 mK. In stark contrast to conventional SIT characterized by a single crossing point [3], the magnetoresistivity isotherms in MTO cross each other in a transition region from 10.2 to 10.6 T (Fig. 2(b)). The crossing points of the magnetoresistivity curves at neighboring temperatures are summarized in the inset of Fig. 2(b), representing the phase boundary between the superconducting and insulating states. To reveal the critical behavior of MTO, the magnetoresistivity isotherms below 600 mK are divided into 10 groups (Fig. S4). Based on the finite size scaling analysis, we derive the effective "critical" exponent $z\nu$ for each group separately [27], here $z$ and $\nu$ are dynamic critical exponent and correlation length exponent, respectively [1,2,74]. As shown in Fig. 2(c), $z\nu$ grows rapidly with increasing field and diverges when approaching the quantum critical point with the characteristic magnetic field $B_c^*$ of 10.588 T. The divergence of $z\nu$ can be well described by the activated scaling law [24] $z\nu \propto (B_c^* - B)^{-\nu\psi}$ with $\nu\psi = 0.33$ (where $\psi$ is the tunneling critical exponent), demonstrating the existence of QGS in 3D MTO superconductors. Moreover, according to previous theoretical literature [75], the divergence of $z\nu$ indicates the violation of the Harris criterion by the clean correlation length exponent $\nu_0$ [76]. Meanwhile, the disorder effect changes the correlation length exponent to the dirty limit value $\nu$, which satisfies $d\nu \geq 2$ [77]. Thus, for 3D systems, $\nu \geq 2/3$ and $\psi$ is predicted to be around 1/2 [24,78]. Based on Harris criterion and the fitting to experimental data, we estimate $\nu\psi = 0.33$ [27]. We notice that the activated scaling law for 1D (red curve, $\nu\psi \approx 1$) and 2D (blue curve, $\nu\psi \approx 0.6$) scenarios significantly deviates from our experimental observation (Fig. S6(a)), further confirming the 3D QGS in MTO. Furthermore, the $\nu\psi$ value for 3D QGS is supported by a recent work reporting QGS in 3D ferromagnet $Ni_{1-x}V_x$ [79]. In spite of different microscopic origins between the SIT and the magnetic QPT, the formation of the superconducting or magnetic rare regions results in similar critical behavior near the infinite randomness quantum critical point, indicating the universality of QGS in 3D systems [27].

Figures 2(d)–(f) reveal the existence of QGS in MTO under parallel magnetic field. The $\rho(T)$ curves at different parallel fields exhibit the SIT behavior (Fig. 2(d)), and the $\rho(B)$ curves at different temperatures reveal a transition region around 10 T (Fig. 2(e)). The crossing points of neighboring $\rho(B)$ curves indicate a larger upper critical field at lower temperatures (the inset of Fig. 2(e)). Based on the finite size scaling analysis (Fig. S5), the divergence of $z\nu$ follows the activated scaling law $z\nu \propto (B_c^* - B)^{-0.33}$ (Fig. 2(f)), which provides solid evidence of QGS in 3D superconductors under parallel magnetic field. The deviation from the 1D (red curve) and 2D (blue curve) theoretical fittings of the activated scaling law (Fig. S6(b)) also confirms the 3D characteristics of QGS. Note that the QGS features under both perpendicular and parallel field are also observed for sample s2 (Fig. S7).

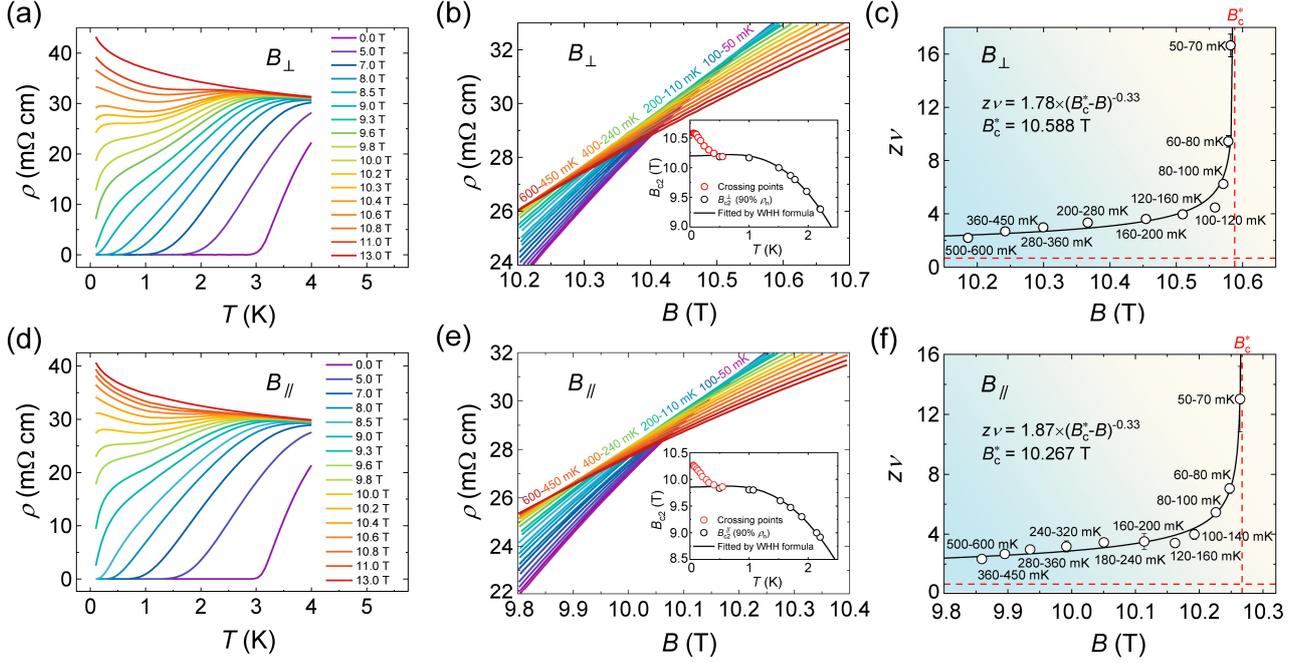

FIG. 2. The quantum Griffiths singularity in MTO (s1). (a) Temperature dependent resistivity $\rho(T)$ under perpendicular fields from 0 to 13 T. (b) Perpendicular field dependent resistivity $\rho(B)$ at temperatures ranging from 50–600 mK. The isotherms were collected with decreasing magnetic field. Inset: Crossing points (red circles) from the magnetoresistivity isotherms. Black circles represent the critical fields defined as the magnetic fields corresponding to 90% $\rho_n$, which are fitted by the WHH formula (black line). (c) Critical exponent $z\nu$ derived from the finite size scaling analysis under perpendicular field. The black line represents the theoretical fitting of the activated scaling law $z\nu \propto (B_c^* - B)^{-0.33}$. The vertical dashed line shows the characteristic magnetic field $B_c^* = 10.588$ T and the horizontal dashed line gives $z\nu = 2/3$. (d)–(f) Similar to (a)–(c) but the data are measured under parallel fields.

In general, the fluctuation effect is more pronounced in low-dimensional superconductors [80] and leads to intriguing QPTs (e.g., SIT and SMT). Figure 3(a) summarizes the strength of fluctuation effect in diverse superconducting systems, including 2D superconducting films and flakes as well as 3D superconducting materials (Table S2). Here the fluctuation strength can be characterized by Ginzburg-Levanyuk parameter $\Delta T/T_c^{onset}$ [81], where the superconducting transition broadening $\Delta T = T_c^{onset} - T_c^{zero}$ at zero field. Note that a small ratio of $B_{c2}^\perp/B_{c2}^{//}$ shows highly anisotropic upper critical field, revealing the characteristic of 2D or layered superconductors.

The 3D superconductivity in MTO is demonstrated by $B_{c2}^\perp/B_{c2}^{//} \approx 1$. As shown in Fig. 3(a) and in Table S2, various 2D superconducting films and flakes (black hollow circles) show relatively large $\Delta T/T_c^{onset}$ ranging from 0.25 to 0.58. In contrast, most bulk superconductors (cyan hollow circles) exhibit comparably small $\Delta T/T_c^{onset}$ between 0 and 0.23. Remarkably, MTO (red solid diamond) exhibits the highest $\Delta T/T_c^{onset}$ of 0.49 among 3D superconductors shown in Fig. 3(a), indicating strong fluctuation effect even comparable to that of 2D superconductors.

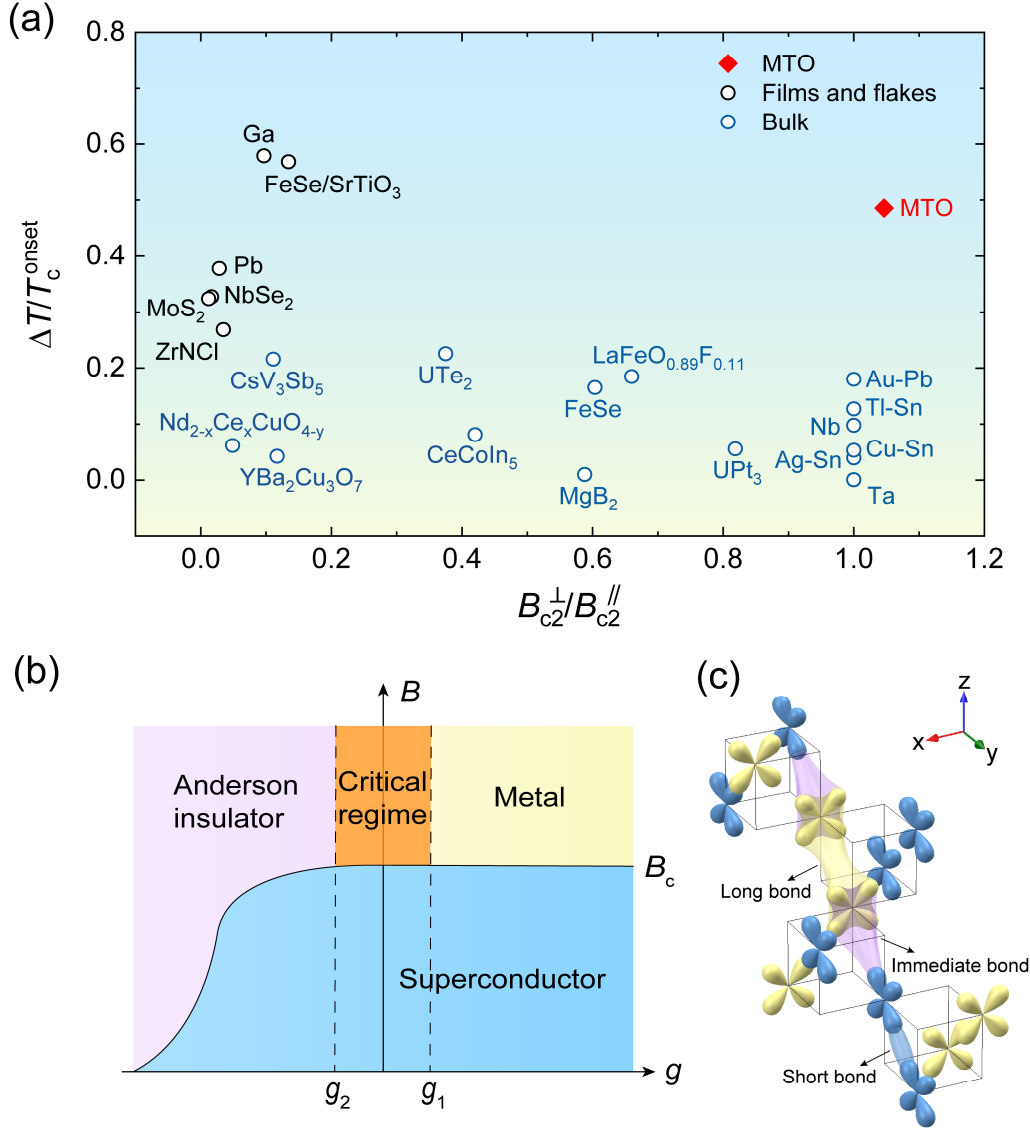

FIG. 3. The disorder enhanced fluctuation effect in MTO. (a) Overview of fluctuation effect in 2D and 3D superconductors. The ratio $\Delta T/T_c^{onset}$ denotes the strength of fluctuation effect and the critical field anisotropy $B_{c2}^{\perp}/B_{c2}^{//}$ indicates the dimension of superconductivity. The black and cyan hollow circles represnent superconducting films or flakes and bulk superconductors, respectively. The red solid diamond represents MTO, whose $\Delta T/T_c^{onset}$ is the largest among 3D superconductors, indicating strong fluctuation effect. (b) Schematic phase diagram of superconducting (blue), metallic (yellow) and insulating (purple) states for 3D systems. The orange region denotes the Anderson critical regime. The horizontal and vertical axes represent the dimensionless conductance $g$ (inversely proportional to disorder strength) and the external magnetic field $B$, respectively. (c) Orbital order in MTO. The $d_{zx}$ and $d_{yz}$ orbitals are shown in yellow and blue, respectively. The long, intermediate, and short bonds are shown in yellow, purple, and blue, respectively.

The extraordinary strong fluctuation effect in 3D superconductor MTO may originate from the Anderson critical regime. Figure 3(b) indicates the schematic phase diagram of superconducting, metallic and Anderson insulating states for 3D systems by varying the dimensionless conductance $g$ (inversely proportional to disorder strength) and the external magnetic field $B$ [81]. With increasing disorder, the Anderson transition occurs when the system changes across the mobility edge from a metallic state to an insulating state. The critical regime ($g_2 < g < g_1$, orange region) is in the vicinity of the mobility edge, sandwiched by the metallic regime ($g > g_1$ and $k_F l > 1$) and the Anderson insulating regime ($g < g_2$ and $k_\text{F}l \ll 1$). The critical regime is of particular interest due to the enhanced fluctuation effect near the mobility edge [9-11]. In this regime, the Anderson localization length exceeds the superconducting coherence length, which enhances the fluctuation of superconducting order parameter even in a 3D system [10]. For 3D superconductor MTO, the small Ioffe-Regel parameter $k_\text{F}l$ (0.28 for perpendicular magnetic field and 0.44 for parallel field) suggests that the normal state of MTO is in this Anderson critical regime (Anderson critical insulator). Moreover, the power-law scaling of the normal state resistivity further confirms the system is in the vicinity of the mobility edge [27,82]. Interestingly, previous theoretical analyses propose that Anderson insulator can exhibit Griffiths phase [83,84], consistent with our experimental observations.

Then we briefly discuss plausible origin of the Anderson critical insulator in MTO. As reported in previous literature, the orbital order in bulk MTO arises from the tetramerization of Ti chains at low temperatures [85,86]. In the electronic band structure, the $d$ orbitals of $Ti^{3+}$ split into the $t_{2g}$ triplet and $e_g$ doublet under crystal field. At low temperatures, the MTO undergoes a cubic to tetragonal structure transition, which further splits the $t_{2g}$ triplet into $d_{zx}/d_{yz}$ bands and $d_{xy}$ band (Fig. S8(a)). Only $d_{zx}/d_{yz}$ bands are favored and occupied by one electron at every Ti site (Fig. 3(c)). The $d$ orbitals of $Ti^{3+}$ form a tetramer structure, i.e., the Ti-Ti bonds are arrayed in order of short, intermediate, long and intermediate bonds (Fig. 3(c)), named as the orbital order. The orbital order opens up a gap for $d_{zx}/d_{yz}$ bands (Fig. S8(b)). With small electron doping, electrons start to occupy the $d_{xy}$ band. The occupied state is close to the bottom of the $d_{xy}$ band (Fig. S8(c)), which is a prerequisite for the Anderson localization. The disorder induced electronic localization in MTO is reminiscent of the possible Anderson localization due to cooperative Jahn-Teller effect in manganites [87-89]. When further increasing electron doping, the system enters the Anderson critical regime (Fig. S8(d)). Moreover, the superconductivity emerges [26,90] when the orbital order is gradually suppressed (Fig. S9). The aforementioned small Ioffe-Regel parameters indicate that the normal state of MTO locates in the Anderson critical regime, and applying magnetic field can induce the QPT between 3D superconductor and the critical insulator. Consequently, the enhanced fluctuation effect in the critical regime enables the observation of 3D QPT with the divergent dynamic critical exponent in MTO.

The strong disorder effect in 3D spinel oxide MTO gives rise to a field-induced 3D superconductor to Anderson critical insulator transition. Within the critical regime of Anderson localization, we observe 3D QGS with divergent critical exponent under both perpendicular and parallel magnetic fields. Our study reveals that strong fluctuation effect in the vicinity of the mobility edge can stabilize the QPT in 3D superconductor and lead to the QGS. Further investigations in correlated systems with orbital orders may explore the exotic quantum criticality near the mobility edge. The experimental investigations along this route may provide a deeper understanding for the quantum dynamics of disordered physical systems.

We thank Jingmin Zhang, Xiaorui Hao, and Jun Xu for STEM characterization of the MTO sample. We thank Kai Liu, Benchao Gong, Dan Shahar, Yingying Peng, Xiquan Zheng, Longxin Pan, and Chengcheng Ji for helpful discussion. This work was financially supported by the National Key Research and Development Program of China (Grant No. 2022YFA1403103), the National Natural Science Foundation of China (Grant No.12488201), the National Key Research and Development Program of China (Grants No. 2018YFA0305604, No. 2023YFA1406500), the Innovation Program for Quantum Science and Technology (2021ZD0302403), the National Natural Science Foundation of China (No. 12174442, No. 12225412, No. 12374037), Beijing National Laboratory for Condensed Matter Physics, CAS Project for Young Scientists in Basic Research (2022YSBR-048), the Center for Materials Genome, Young Elite Scientists Sponsorship Program by CAST (No. 2023QNRC001), Young Elite Scientists Sponsorship Program by BAST (No. BYESS2023452), and the Fundamental Research Funds for the Central Universities.

S.Q., Y.L., Z.W. and F.C. contributed equally to this work.

*Corresponding author.

Jian Wang (jianwangphysics@pku.edu.cn)

†Corresponding author.

Kui Jin (kuijin@iphy.ac.cn)

‡Corresponding author.

Haiwen Liu (haiwen.liu@bnu.edu.cn)

Supplemental Material for

# Quantum Griffiths Singularity in a Three-Dimensional Superconductor to Anderson Critical Insulator Transition

Shichao Qi,[1] Yi Liu,[2,3] Ziqiao Wang,[1] Fucong Chen,[4] Qian Li,[4] Haoran Ji,[1] Rao Li,[2] Yanan Li,[1] Jingchao Fang,[1] Haiwen Liu,[5,6,‡] Fa Wang,[1] Kui Jin,[4,7,8,†] X. C. Xie,[1,6,9] and Jian Wang[1,9,10,*]

[1]*International Center for Quantum Materials, School of Physics, Peking University, Beijing 100871, China*
[2] *Department of Physics and Beijing Key Laboratory of Opto-electronic Functional Materials & Micro-nano Devices, Renmin University of China, Beijing 100872, China*
[3]*Key Laboratory of Quantum State Construction and Manipulation (Ministry of Education), Renmin University of China, Beijing 100872, China*
[4]*Beijing National Laboratory for Condensed Matter Physics, Institute of Physics, Chinese Academy of Sciences, Beijing 100190, China*
[5]*Center for Advanced Quantum Studies, Department of Physics, Beijing Normal University, Beijing 100875, China*
[6]*Interdisciplinary Center for Theoretical Physics and Information Sciences, Fudan University, Shanghai 200433, China*
[7]*School of Physical Sciences, University of Chinese Academy of Sciences, Beijing 100049, China*
[8]*Songshan Lake Materials Laboratory, Dongguan, Guangdong 523808, China*
[9]*Hefei National Laboratory, Hefei 230088, China*
[10]*Collaborative Innovation Center of Quantum Matter, Beijing 100871, China*

S.Q., Y.L., Z.W. and F.C. contributed equally to this work.
[*]Corresponding author.
Jian Wang (jianwangphysics@pku.edu.cn)
[†]Corresponding author.
Kui Jin (kuijin@iphy.ac.cn)
[‡]Corresponding author.
Haiwen Liu (haiwen.liu@bnu.edu.cn)

## Contents

## I. Methods

**Sample preparation and characterization.** The crystalline MTO films (377–474 nm) were grown on the (00*l*)-oriented cubic spinel $MgAl_2O_4$ (MAO) insulating substrates via pulsed laser deposition. The pulse energy was about 250 mJ and the repetition rate was 10 Hz. The deposition temperature was set to be 800 °C. The chamber vacuum was maintained under $10^{-6}$ torr. The X-ray diffraction characterization was carried out in Rigaku Smartlab 9kW at room temperature, using Cu Kα radiation. The X-ray diffraction result confirms that the (00$l$) lattice plane of MTO films is well parallel to that of MAO substrates. The high-resolution scanning transmission electron microscopy images were observed in a Titan Cubed Themis G2 double Cs-corrected scanning transmission electron microscope.

**Transport measurements.** Standard four-electrode method (inset of Fig. 1(b)) was used for ex situ transport measurements. Two indium strips along the width of the sample served as the current electrodes (I+ and I-), making the current homogenously pass through the sample. The other two indium electrodes (V+ and V-) were pressed in the middle of the sample to measure the longitudinal voltage. The transport measurements were carried out in a commercial physical property measurement system (Quantum Design, PPMS-16 with the dilution refrigerator and rotator options). The excitation current was perpendicular to the magnetic field during the magnetoresistivity measurements. All the isotherms in Figs. 2, S4, S5 and S7 were collected with decreasing magnetic field.

## II. The magnetic critical fields fitted by WHH formula

The perpendicular and parallel magnetic critical fields (Fig. 1(e)) are fitted by WHH formula $ln\frac{1}{t} = \left(\frac{1}{2} + \frac{i\lambda_{SO}}{4\gamma}\right)\psi\left(\frac{1}{2} + \frac{\bar{h}+\frac{1}{2}\lambda_{SO}+i\gamma}{2t}\right) + \left(\frac{1}{2} - \frac{i\lambda_{SO}}{4\gamma}\right)\psi\left(\frac{1}{2} + \frac{\bar{h}+\frac{1}{2}\lambda_{SO}-i\gamma}{2t}\right) - \psi\left(\frac{1}{2}\right)$ [1]. Here, $\gamma = [(\alpha\bar{h})^2 - (\frac{1}{2}\lambda_{SO})^2]^{1/2}$, $t = \frac{T}{T_c}$, $\lambda_{SO}$ represents the strength of spin-orbit interaction, $\alpha$ is the strength of spin paramagnetic effect, $\bar{h}$ denotes the strength of orbital effect and $\psi(x)$ is the digamma function. The fitting shows $\alpha = 5.29\ (3.41)$ and the Ioffe-Regel (IR) parameter $k_F l = 0.28\ (0.44)$ for perpendicular (parallel) field, indicating the large spin pair-breaking effect in MTO. The difference of IR parameters derived from perpendicular and parallel critical fields may originate from the anisotropic Fermi surface. With increasing the electron doping, the $d_{xy}$ band is occupied and the $d_{zx}$-$d_{yz}$ orbital order is gradually suppressed as shown in Figs. S8 and S9. The density of states for the $d_{xy}$ band becomes larger than that for $d_{zx}$ and $d_{yz}$ bands [2], which implies an anisotropic Fermi surface and different Fermi wave vector $k_F$ for the perpendicular and parallel directions.

## III. Discussions on the origin of disorder in MTO

The disorder may originate from the lattice distortions of the MTO film. Although the MTO film shows crystalline lattice structure, the high-resolution STEM images show signatures of lattice distortions in the circulated region (Fig. S1(e)). Another possible origin of disorder is the atom vacancies in MTO. As previously reported, the superconductivity is realized in the MTO film via reducing the ratio of Mg/Ti which increases the electron doping [2]. The reduction of the Mg/Ti ratio produces atom vacancies, which also increases the disorder strength of MTO. Besides, the oxygen atom vacancies may exist because the sample is grown in a vacuum chamber with pressure under $10^{-6}$ torr.

## IV. The fermionic nature of the insulating state

When the perpendicular magnetic field increases, the MTO undergoes a superconductor to insulator transition. As shown in Fig. S3(a), the MTO (s1) enters an insulating state when the magnetic field reaches 13 T. The red dashed line in Fig. S3(a) is the fitting curves with superinsulator (bosonic insulator) formula [3], written as $\rho = \rho_0 * \exp\left(A * \exp\sqrt{\frac{b}{(T/T_{CBKT})-1}}\right)$, where $\rho_0$, $A$ and $b$ are material-dependent parameters and $T_{CBKT}$ is the charge-BKT transition temperature. The experimental data deviates greatly from the fitting curves and the resistivity increases more slowly than bosonic insulator prediction with decreasing temperature, indicating the insulating phase is dominated by quasi-particles rather than incoherent Cooper pairs. Besides, the slope of the insulating state for sample s2 increases smoothly as cooling down from 300 K to 2 K under 15 T magnetic field (Fig. S3(b)). The upward trend below $T_c^{\text{onset}}(0\text{ T})$ is similar to that of the normal state, indicating the fermionic nature of the insulating state.

## V. The finite size scaling analysis for 3D QGS state

In general, for a conventional QPT (e.g. SIT and SMT) in 2D superconductors [4], the resistance can be expressed as $R = R_c \cdot \Phi\{|B - B_c|T^{-1/z\nu}\}$, where $\Phi$ is an arbitrary function with $\Phi(0) = 1$, $R_c$ and $B_c$ are the critical resistance and critical magnetic field when approaching the quantum critical point. As for 3D SIT/SMT, the resistivity of the system satisfies $\rho = \rho_c \cdot \tilde{L} \cdot \Phi\{|B - B_c|T^{-1/z\nu}\}$, where $\rho_c$ is the critical resistivity, the dimensionless thermal length $\tilde{L}$ is defined as $(\frac{T^*}{T})^{1/z}$ and $T^*$ is the characteristic temperature associated with the QPT. The temperature dependence of dimensionless thermal length $\tilde{L}$ is taken from the general consideration of the critical resistivity near quantum critical points [5,6]. Moreover, in the quantum Griffiths phase, the dynamics is slowed down in presence of quenched disorder [7]. Then the resistivity can be described as $\rho = \rho_c \cdot \tilde{L} \cdot \Phi\{|B - B_c|[\ln(T^*/T)]^{(1/\nu\psi)}\}$ where $\tilde{L} = [\ln(T^*/T)]^p$ and $p$ is the power index dependent on materials. The finite size scaling analysis is normally utilized to determine the effective "critical" exponents $z\nu$ of the QPT. The magnetoresistivity isotherms of MTO at different temperatures cross each other in a transition region with multiple crossing points. To obtain the effective "critical" exponents $z\nu$, we divide magnetoresistivity isotherms into several groups and the crossing region for each group can be effectively regarded as one crossing point. Here we approximately treat the $\tilde{L} = [\ln(T^*/T)]^p$ as a constant, since the temperature regime for each group is very small. Thus, the resistivity follows the scaling formula $\rho(B,T)/\rho_c = \Phi(|B - B_c|t)$, where $t \equiv (T/T_0)^{-1/z\nu}$ and $T_0$ is the lowest temperature of this group. The parameter $t$ is determined by rescaling the normalized resistivity curves as a function of $|B - B_c|t$ to match that of the lowest temperature $T_0$. Then the effective "critical" exponent $z\nu$ can be extracted by the linear fitting between $\ln(T/T_0)$ and $\ln(t)$. The scaling results are summarized in Figs. S4 and S5.

## VI. Discussions on Harris criterion and estimation of critical exponent

The Harris criterion was proposed to determine the stability of the clean critical point against the disorder [8]. If the clean correlation length critical exponent $\nu_0$ and the dimensionality $d$ fulfill the Harris criterion $d\nu_0 \geq 2$, the critical behavior is basically unaffected by the disorder. On the contrary, if the critical exponent violates the Harris criterion ($d\nu_0 < 2$), the critical behavior is destabilized by the disorder. In this case, the value of clean correlation length critical exponent $\nu_0$ is significantly influenced by the disorder and changes to a new value $\nu$, which satisfies $d\nu \geq 2$ [9,10].

Theoretically, the emergence of QGS (i.e., the divergence of the dynamical critical exponent $z$ when approaching the quantum critical point) requires the violation of Harris criterion [11]. Otherwise, the dynamical critical exponent $z$ remains finite. In our experiment, for the disordered superconducting MTO, $z\nu$ increases significantly and then diverges when approaching the quantum critical point and does not have a trend to saturate down to the millikelvin range. Although the clean correlation length critical exponent $\nu_0$ cannot be detected experimentally, the divergence of $z\nu$ indicates the violation of the Harris criterion ($d\nu_0 < 2$).

Based on the aforementioned discussions, we roughly estimated the dirty correlation length exponent $\nu \geq 2/3$ based on $d\nu \geq 2$ and $d = 3$. Considering the tunneling critical exponent $\psi=0.5$ in 1D systems and varies very weakly with dimensionality [12,13], the $\nu\psi$ is estimated to fulfill $\nu\psi \geq 0.33$. As shown in Figs. S10(a) and S10(b), the best fitting of the divergent $z\nu$ yields the fitting parameter $\nu\psi = 0.33$ under both perpendicular and parallel magnetic fields.

Compared to the 1D and 2D cases, it's difficult to calculate the critical exponent $\nu\psi$ in 3D systems [12-14]. We notice that a numerical renormalization group method study deduces $\nu \approx 0.97$ and $\psi \approx 0.46$ for 3D random transverse-field Ising model, yielding $\nu\psi \approx 0.45$ [10]. The fitting results of the divergent $z\nu$ with $\nu\psi \approx 0.45$ are shown in Figs. S10(c) and (d), which have larger deviation from the experimental results compared to the fitting curves using $\nu\psi \approx 0.33$. The approaching to an Anderson critical insulator may influence the value of $\nu$ in our system [15], leading to a different value compared to the 3D random transverse-field Ising model. Moreover, we notice a recent work reporting QGS in 3D ferromagnet $Ni_{1-x}V_x$ whose $\nu\psi = 0.34$ [16]. These results indicate that in 3D systems with QGS, the value of correlation length critical exponent $\nu$ may deviate from the prediction of 3D random transverse field Ising model.

### VII. The power-law scaling of the normal state of MTO

Theoretically, a system in the quantum critical regime of the Anderson insulator–metal transition should exhibit power-law scaling of its resistivity: $\rho = c * T^{-\beta}$, where $c$ is the material dependent coefficient and $\beta$ is the power-law exponent [17]. We have performed the power-law scaling of the normal state for the MTO film (sample s1) as shown in Fig. S11(a). The fitted power-law exponent $\beta$ (0.215) is smaller than 1/3, indicating that sample s1 is in the quantum critical regime of the Anderson insulator-to-metal transition but still in the metallic side. Moreover, we have also performed the power-law scaling for MTO samples (s3 and s4) which has lower electron doping (Fig. S11(b) and (c)). The power-law fitting of the normal state $\rho(T)$ below 25 K yields $\beta = 0.359$ for s3 and $\beta = 0.563$ for s4. The analysis of three samples indicates that the MTO films evolve from the metallic to the insulating side of the Anderson transition with decreasing the electron doping. The power-law scaling for of the normal states, in combination with the estimations of the Ioffe-Regel parameter, strongly supports that the superconducting MTO sample (s1) is near the mobility edge. Interestingly, previous theoretical works propose that Anderson insulator can exhibit Griffiths phase [15,18], which is well consistent with our experimental results.

### VIII. Discussions on the kink behavior of $\rho(T)$ curves under high magnetic fields

The kink behavior in the $\rho(T)$ curves under high fields (Figs. 2(a) and 2(d)) may arise from the competition between superconductivity and the residual orbital order. As reported previously, the orbital order forms in bulk MTO which shows the insulating behavior as the temperature decreases [19,20]. For the MTO films, the orbital order is gradually suppressed and the superconductivity emerges with increasing the electron doping [2]. As shown in Figs. 2(a) and 2(d), at zero field, the superconductivity dominates the transport properties of the MTO film and gives rise to the smooth transition. When the superconductivity is largely weakened under high magnetic fields, the competition between the superconducting order and the residual orbital order could give rise to the kink behavior of the $\rho(T)$ curves.

### IX. Discussions on the observation of QGS in superconducting systems

So far, the divergence of critical exponent $z\nu$ approaching the quantum critical point has been revealed in a wide variety of 2D superconducting systems [21-23]. Particularly, the critical behavior $z\nu \propto |B - B_c^*|^{-0.6}$ (where $B_c^*$ is the quantum critical point) has been confirmed in diverse 2D superconducting systems. The universal critical behavior can be well explained in the framework of QGS, and to the best of our knowledge, the QGS is the only model to quantitatively explain the divergence of the critical exponent in 2D superconductors.

Theoretically, the quantum Griffiths singularity can exist in superconducting systems. The QGS was initially proposed by D. Fisher in random transverse-field Ising model (RTFIM) [12,24]. Then T. Vojta and collaborators showed that, the O(*N*) Landau-Ginzburg-Wilson order-parameter field theory with $N \geq 2$ and Ohmic dissipation is in the same universality class as the transverse-field Ising model [25-28]. These theoretical studies lay the theoretical foundation for QGS in superconductors, represented by O(2) Landau-Ginzburg-Wilson theory with Ohmic dissipation. In the vicinity of infinite-randomness quantum critical point, the slow dynamics of rare regions (the temporal correlation length $\xi_\tau$ follows an activated scaling $ln(\xi_\tau) \propto L^\psi$, where $L$ is the size of the rare region and $\psi$ is the tunneling critical exponent) dominates the critical behavior of global system and gives rise to divergent effective critical exponent $zv \propto |B - B_c^*|^{-\nu\psi}$ with $\nu$ denoting the correlation critical exponent [12,24,26,27], which provides a physical understanding of QGS.

Previously it is argued [29,30] that the prerequisite for a true QGS is that the susceptibility of the rare region needs to diverge exponentially with its volume (or area in 2D). Conversely, the coupling to the heat bath grows in proportion to the surface or edge of the rare region with a limited superconducting coherence length $\xi$. From our points of view, this argument does not apply to the region close to the phase boundary. The rare region is composed of $N$ superconducting grains which are coupled via the short-range Josephson coupling, and the size is $R_i$ ($i = 1,2,...,N$) for each grain. The total volume of rare region ($V$) is approximately the sum of the grain volume $V \approx \sum_i R_i^d$, which increases to infinite ($V \to \infty$) when approaching the quantum critical point. The probability of finding such rare region falls exponentially with its volume and the parameter $c$ ($P_{RR} \sim \exp(-cV)$), here $c$ is propotional to the distance to the critical point [9]. Along the phase boundary of superconductor–metal transition (or superconductor to Anderson critical insulator transition), the coherence length $\xi$ is slightly smaller than the grain size $R_i$ ($|\xi - R_i| \ll R_i$ for near critical grains) [29]. Thus, the coupling between the superconducting rare region and fermionic heat bath occurs in a region ($\tilde{V} \approx \sum_i R_i^{d-1}\xi$) that is very close to the volume of rare region $V \approx \sum_i R_i^d$, which means the coupling to the heat bath could penetrate almost the whole rare region. Furthermore, approaching the quantum critical point of the 3D superconductor to Anderson critical insulator transition, the multifractality of electron wavefunctions [31] may further enhance the coupling between the superconducting rare regions and the fermionic heat bath. The susceptibility of individual rare region grows exponentially with the volume of the coupling to the heat bath ($\chi_{RR} \sim \exp(a\tilde{V})$) [9], in which the parameter $a$ is propotional to the dimensionless inhomogeneity strength and does not change when tuning the phase transition parameter (for example in RTFIM $a \propto \sigma(J)/\langle J \rangle$ with $\sigma(J)$ denoting the standard deviation of random Ising coupling $J$ and $\langle J \rangle$ denoting the mean value of random $J$). Consequently, the susceptibility of the whole system with lots of near critical grains reads $\chi = \chi_{RR} \cdot P_{RR} \sim \exp(a\tilde{V})\exp(-cV) \sim \exp((a\xi/\langle R \rangle - c)V)$, with $\langle R \rangle \equiv \sum_i R_i^d / \sum_i R_i^{d-1}$ and $\xi/\langle R \rangle \approx 1$. When approaching the quantum critical point, $c$ approaches zero but $a$ is still finite [9], thus the susceptibility of the system diverges near the critical point. Taken together, the prerequisite for QGS can be satisfied in most experimentally achievable ultralow temperature regime.

In the following, we clarify that the MTO shows the quantum Griffiths singularity down to low temperatures instead of a smeared phase transition. Based on the experimental observations of quantum phase transition in superconducting MTO and the related theoretical models, we conclude that the MTO system fulfills Class B rather than Class C in ref. [9]. Firstly, from the experimental aspect, our observation in superconducting transition of MTO gives the divergent critical exponent $zv$ with the temperature down to 50 mK, in sharp contrast to the smeared quantum phase transition of Class C ($zv$ increases to infinite value at a relatively high temperature and does not follow the activated scaling law at lower temperatures). Secondly, from the theoretical perspective, the quantum phase transition in our system is represented by a disordered O(2) Landau-Ginzburg-Wilson theory with the Ohmic dissipation, which is identical to the random-transverse field Ising model without Ohmic dissipation and belongs to the QGS universality (Class B) [25-28]. In-depth theoretical discussions on these points are summarized in Sec. 4.2 and 4.3 of ref. [9]. Lastly, it is proposed that, the smeared phase transition would occur in the presence of long-range Josephson coupling between superconducting grains [28]. However, in the magnetic-field-induced superconducting quantum phase transition, the Josephson coupling has a short characteristic length under the magnetic field [29]. Thus, the smeared phase transition is unlikely to occur in the magnetic-field-induced quantum phase transition of MTO with short-range Josephson coupling.

# X. Figures and Tables

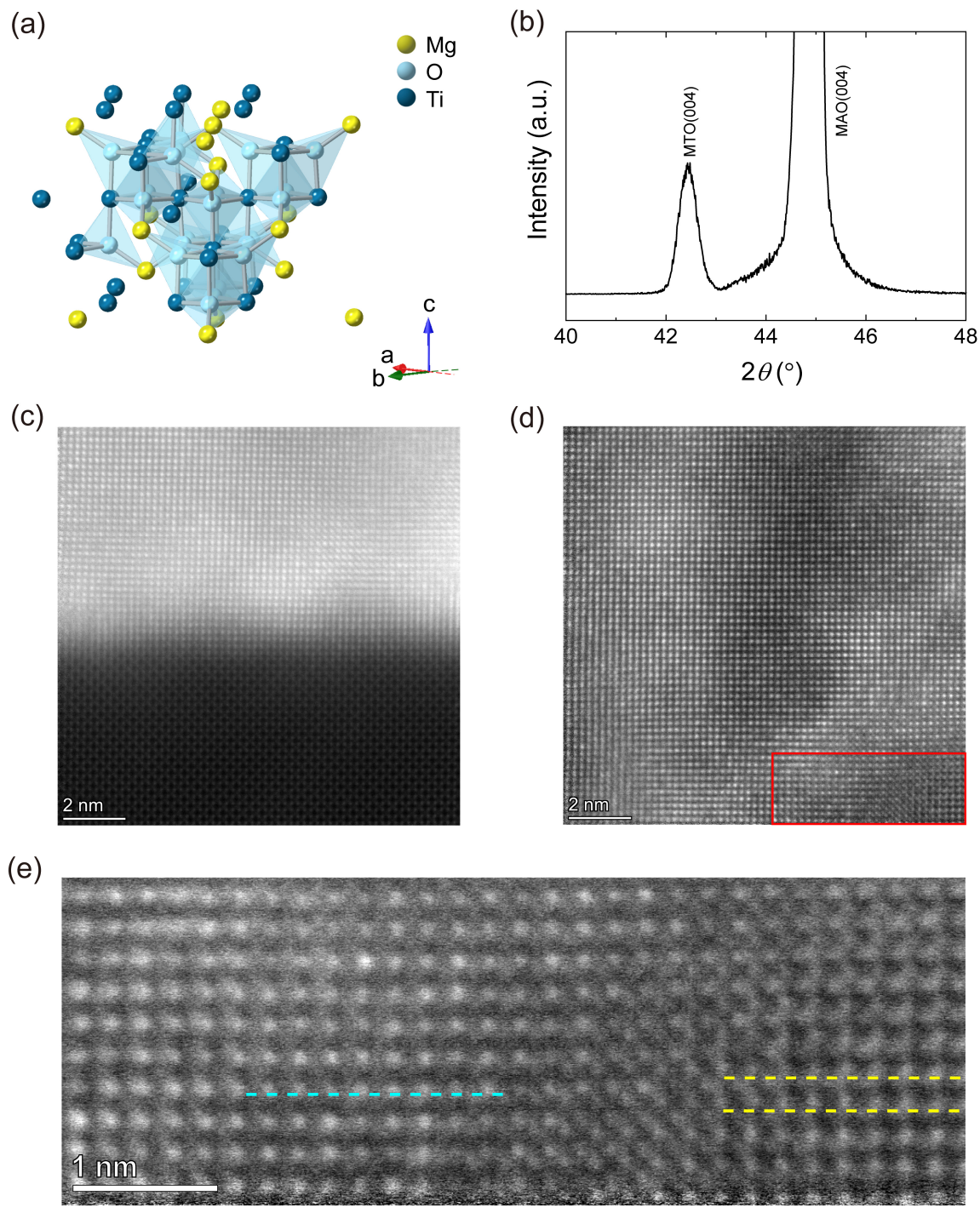

FIG. S1. Structure characterization of MTO. (a) The schematic of the spinel crystal structure of MTO. (b) The X-ray diffraction spectra of MTO. (c) The high-resolution scanning transmission electron microscopy (STEM) image of the interface between MAO and MTO. The brighter atomic structure forms MTO lattice. (d) The high-resolution STEM image of MTO. (e) The zoom-in image of the enclosed region in (d). The blue and yellow dashed lines represent different rows of atoms.

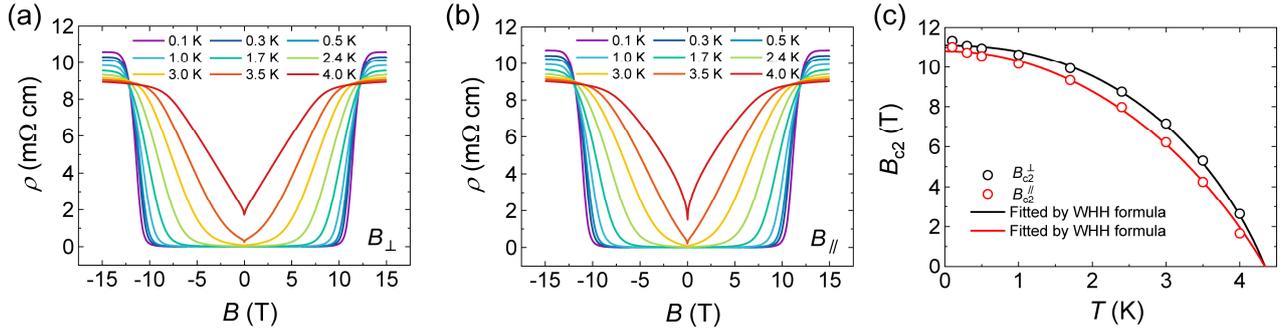

FIG. S2. 3D superconductivity in MTO (sample s2). (a) (b) Magnetoresistivity isotherms of s2 under perpendicular (a) and parallel (b) magnetic fields, respectively. (c) Temperature dependence of perpendicular (black spheres) and parallel (red spheres) critical fields (defined as the magnetic field corresponding to 50% of the normal state resistivity) for sample s2. The solid lines are theoretical fittings based on the WHH formula.

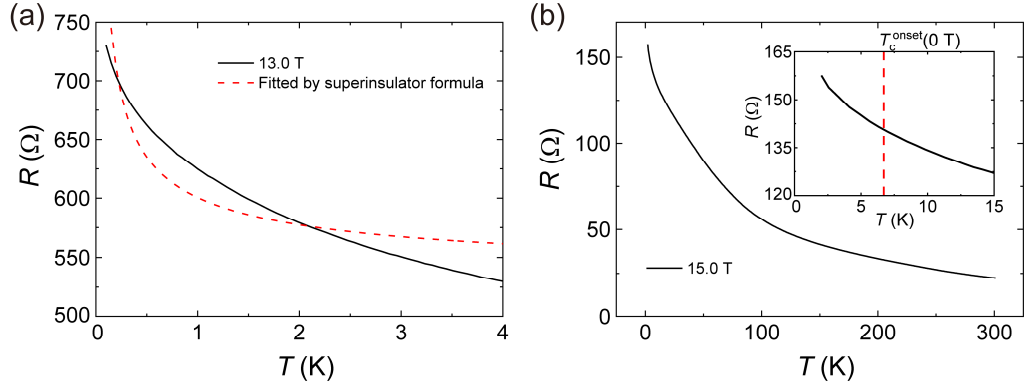

FIG. S3. The fermionic nature of the insulating state of MTO. (a) Temperature dependence of resistance for sample s1 under perpendicular magnetic field 13 T. The red dashed line is the fitting curve based on superinsulator (bosonic insulator) formula. (b) Temperature dependence of resistance for sample s2 from 2 K to 300 K under perpendicular magnetic field 15 T. The inset is an enlarged view of the $R(T)$ curve from 2 K to 15 K. The red dashed line denotes the critical temperature $T_c^{onset}$ of superconductivity for sample s2 under zero magnetic field.

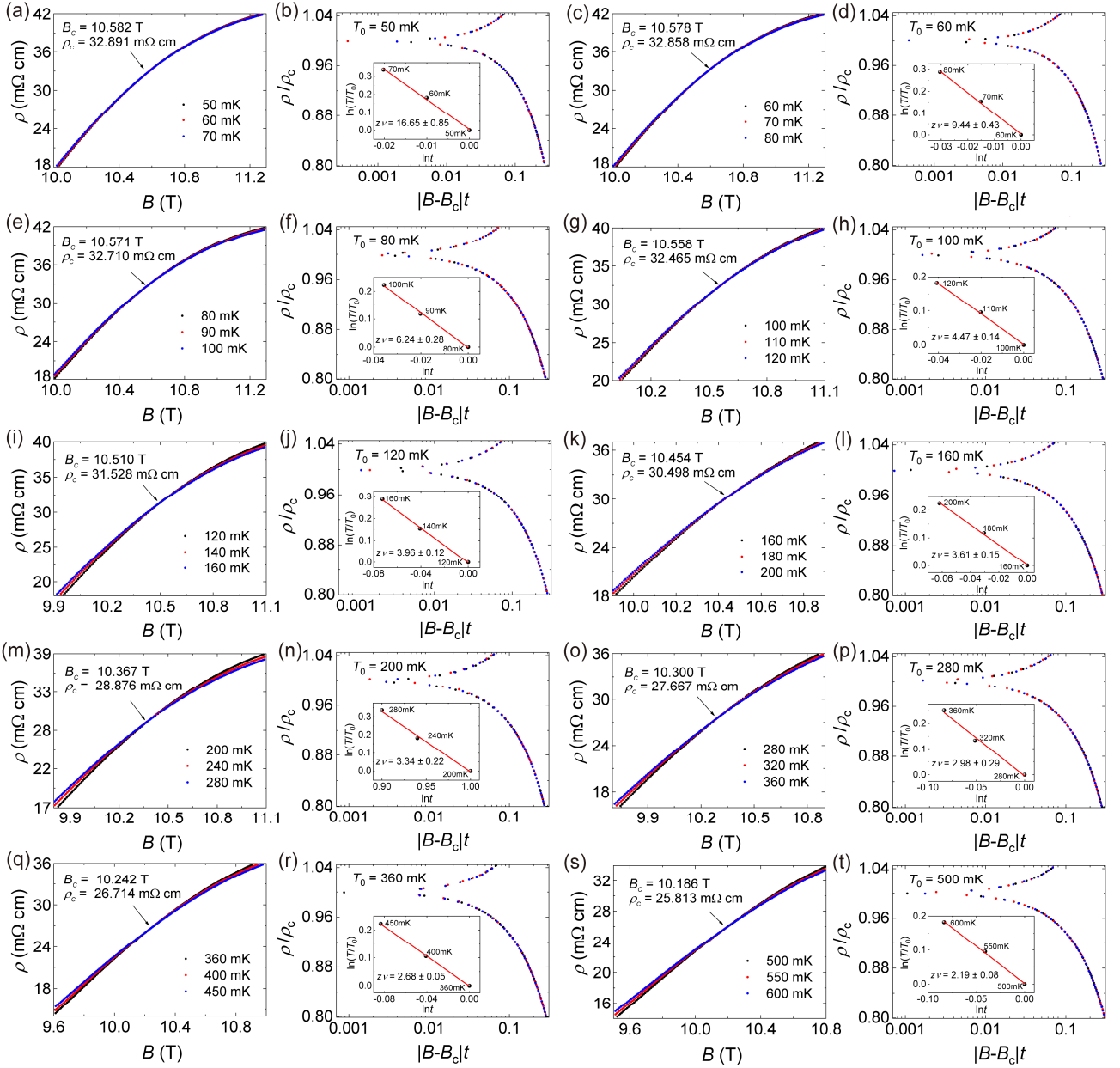

FIG. S4. Finite size scaling analysis for the MTO (s1) under perpendicular magnetic field at temperatures from 50 to 600 mK. (a) (c) (e) (g) (i) (k) (m) (o) (q) (s) Perpendicular magnetic field dependence of resistivity in different temperature regime. (b) (d) (f) (h) (j) (l) (n) (p) (r) (t) Corresponding normalized resistivity as a function of scaling variable $|B - B_c|t$, with $t = (T/T_0)^{-1/z\nu}$. Inset: linear fitting between $\ln(T/T_0)$ and $\ln t$ gives effective "critical" exponent $z\nu$.

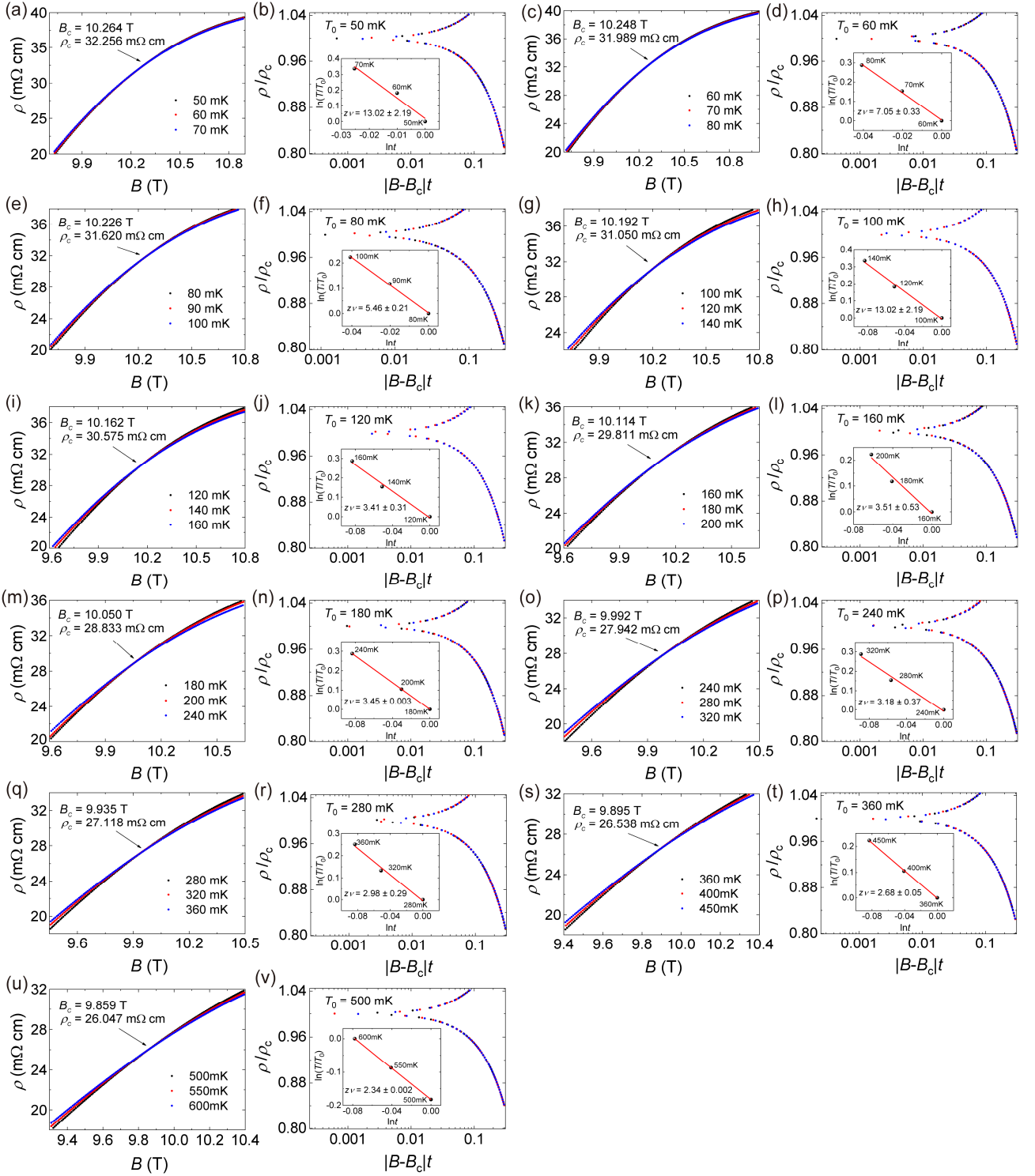

FIG. S5. Finite size scaling analysis for the MTO (s1) under parallel magnetic field at temperatures from 50 to 600 mK. (a) (c) (e) (g) (i) (k) (m) (o) (q) (s) (u) Parallel magnetic field dependence of resistivity in different temperature regime. (b) (d) (f) (h) (j) (l) (n) (p) (r) (t) (v) Corresponding normalized resistivity as a function of scaling variable $|B - B_c|t$, with $t = (T/T_0)^{-1/z\nu}$. Inset: linear fitting between $\ln(T/T_0)$ and $\ln t$ gives effective "critical" exponent $z\nu$.

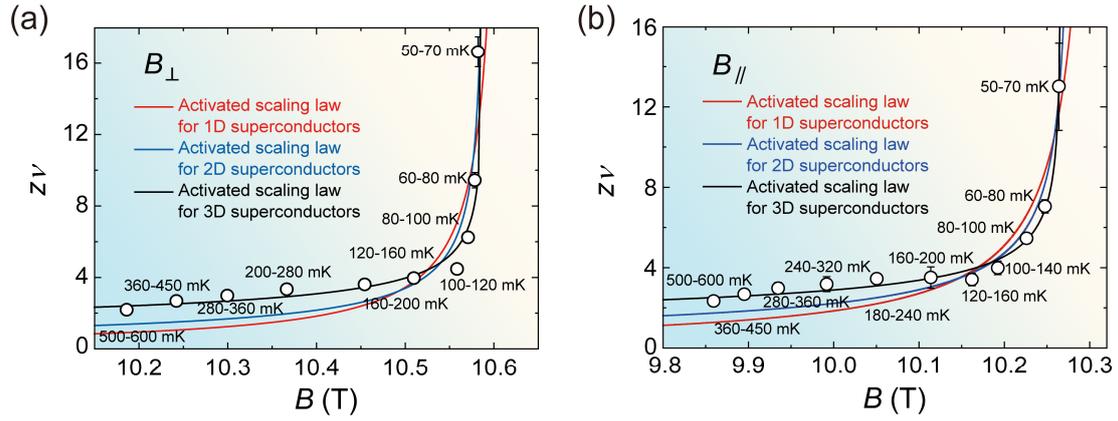

FIG. S6. Evidence of 3D QGS in MTO (s1). (a) (b) The divergence of critical exponent $z\nu$ under perpendicular (a) and parallel (b) magnetic fields. The solid black line is the theoretical fitting based on the activated scaling law for 3D superconductors with $\nu\psi = 0.33$. The red and blue solid lines represent the activated scaling law for 1D ($\nu\psi = 1$) and 2D ($\nu\psi = 0.6$) scenarios, deviating from the experimental observation.

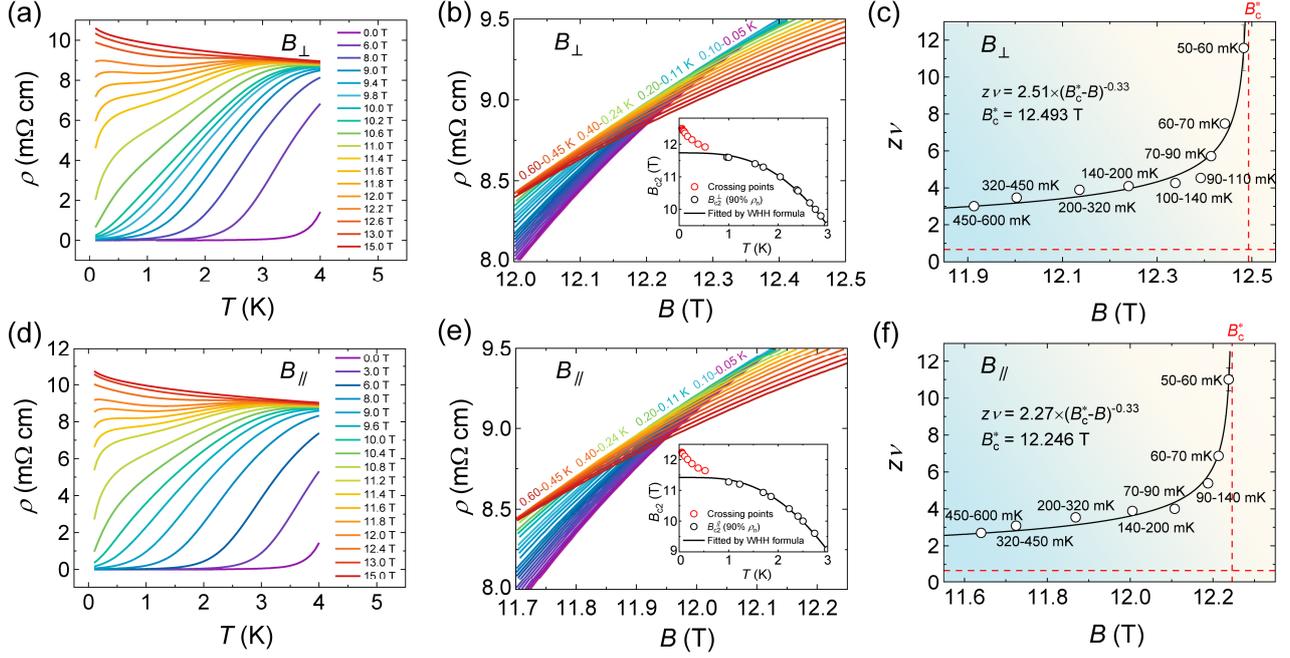

FIG. S7. The QGS in another MTO sample (s2). (a) Temperature dependence of resistivity $\rho$ under perpendicular magnetic fields from 0 to 15 T. (b) Perpendicular magnetic field dependence of resistivity $\rho$ at different temperatures ranging from 50–600 mK. Crossing points from the magnetoresistivity isotherms are shown in the inset. (c) Critical exponent $z\nu$ derived from the finite size scaling analysis under perpendicular field. The solid black line is the theoretical fitting based on the activated scaling law. The vertical dashed line shows the fitted critical magnetic field $B_c^* = 12.493$ T and the horizontal dashed line gives $z\nu = 2/3$. (d)–(f) Similar to (a)–(c) but the data are measured when the magnetic field is parallel to the ab-plane of MTO.

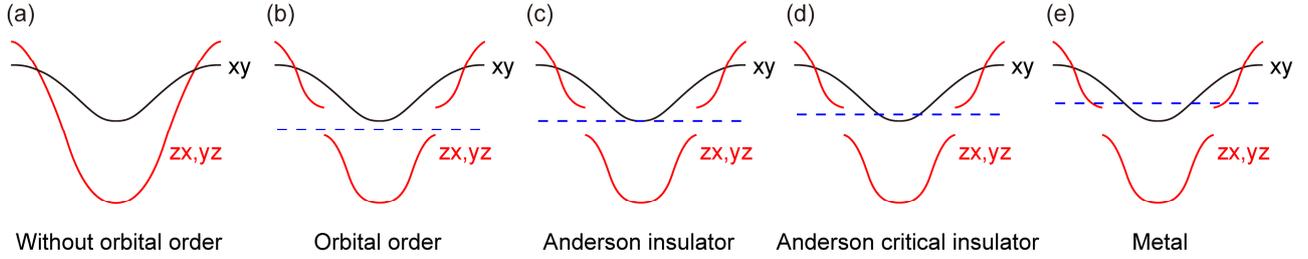

FIG. S8. Schematic of electronic band structure in MTO. (a) Schematic of electronic band structure without considering orbital order. The black curve represents $d_{xy}$ band and the red curve denotes the $d_{zx}$ and $d_{yz}$ bands. (b) The orbital order opens up a gap for the $d_{zx}$ and $d_{yz}$ bands. The blue dashed line represents the Fermi surface. (c) With small electron doping, the occupied state is close to the bottom of the $d_{xy}$ band and the system is an Anderson insulator. (d) With increasing electron doping, the system enters the Anderson critical regime (i.e., Anderson critical insulator). (e) The system enters the metallic state with further increase of electron doping.

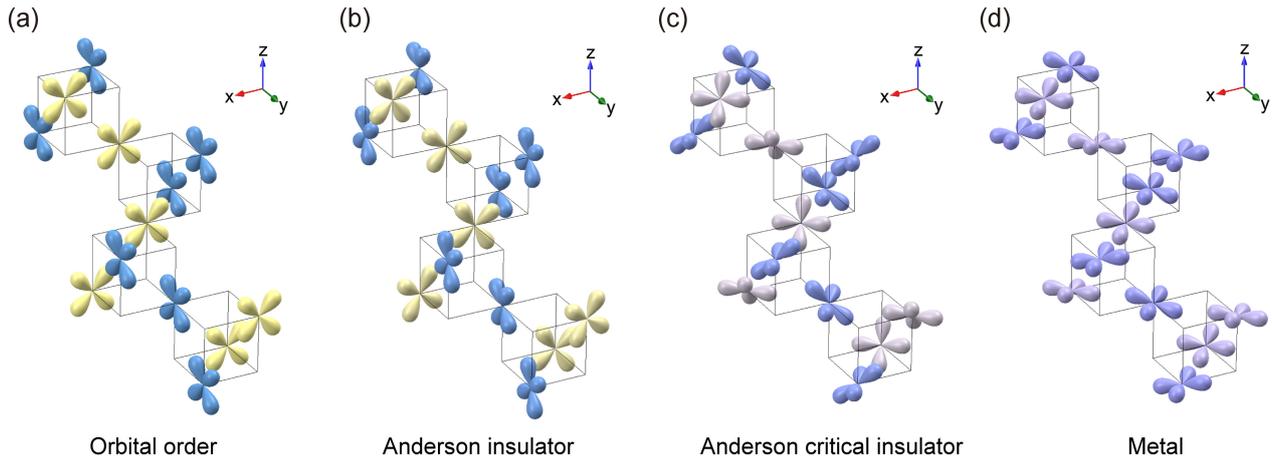

FIG. S9. Orbital occupation in MTO with increasing electron doping. (a) The $d_{zx}$ and $d_{yz}$ bands are occupied by one electron at every Ti site and form the orbital order in MTO. (b)–(d) With increasing electron doping, the electrons start to occupy the $d_{xy}$ band and the $d_{zx}$-$d_{yz}$ orbital order is gradually suppressed. The system evolves into the Anderson insulating (b), the Anderson critical insulating (c) and the metallic (d) states in turn.

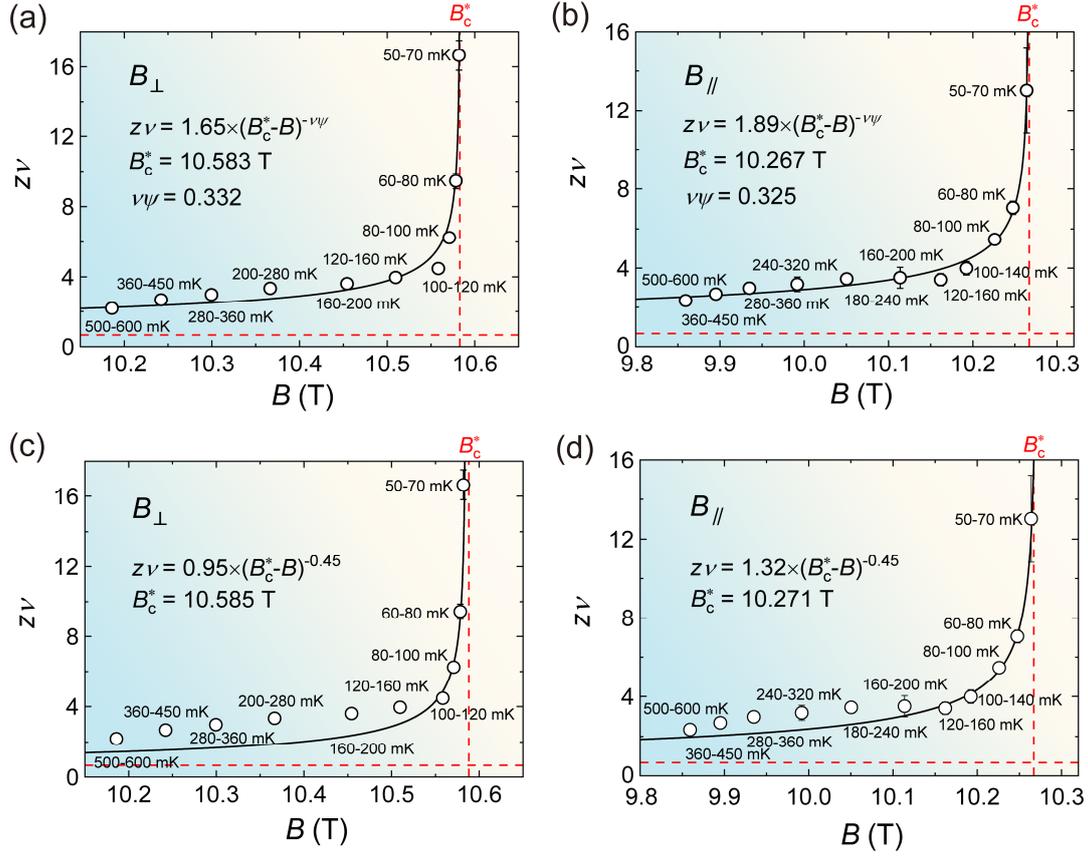

FIG. S10. Critical exponent $z\nu$ derived from the finite size scaling analysis under perpendicular (a) (c) and parallel (b) (d) magnetic fields. (a) (b) The solid black curve represents the theoretical fitting of the activated scaling law with free parameter $\nu\psi$. (c) (d) The solid black curve represents the theoretical fitting of the activated scaling law with parameter $\nu\psi = 0.45$.

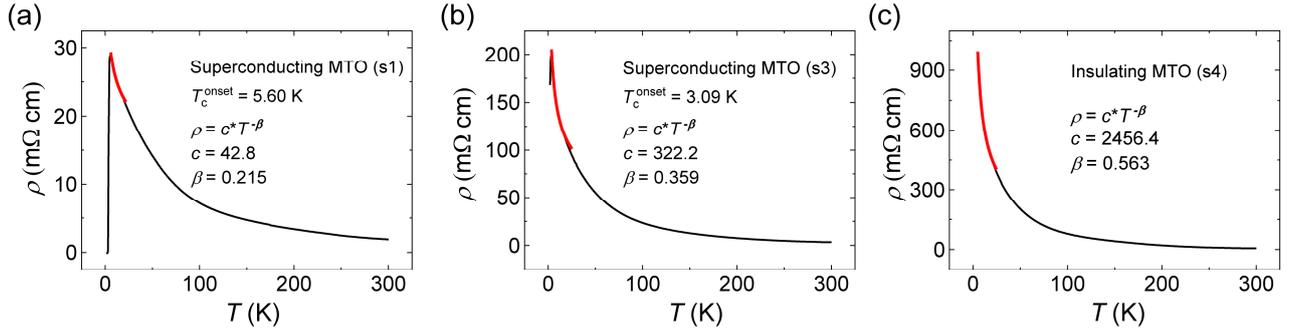

FIG. S11. The power-law scaling analysis of the normal state of MTO under zero magnetic field. (a) The $\rho(T)$ curve of the superconducting MTO film (s1) with $T_c^{onset} = 5.60$ K. The red line represents the power-law fitting with parameter $\beta = 0.215$. (b) The $\rho(T)$ curve of the superconducting MTO film (s3) with $T_c^{onset} = 3.09$ K. The red line represents the power-law fitting with parameter $\beta = 0.359$. (c) The $\rho(T)$ curve of the insulating MTO film (s4). The red line represents the power-law fitting with parameter $\beta = 0.563$. These scaling analyses confirm that the samples locate near the mobility edge.

Table S1. The Ginzburg-Landau coherence length and the thickness for MTO and typical 2D superconductors.

| Material | Thickness $d$ (nm) | Coherence length $\xi_{GL}$ (nm) | $d/\xi_{GL}$ |
|---|---|---|---|
| MgTi$_2$O$_4$ | 377 | 2.85 | 132 |
| Pb film [32] | 1.12 | 12.7 | 0.0882 |
| Ga film [33] | 0.552 | 14.3 | 0.0386 |
| NbSe$_2$ flake [34] | 1.2 | 16.7 | 0.0719 |
| FeSe/SrTiO$_3$ [35] | 0.55 | 2.46 | 0.224 |

Table S2. The fluctuations and anisotropy of various superconductors in Fig. 3(a). The anisotropy of critical magnetic field $B_{c2}^{\perp}/B_{c2}^{//}$ of bulk superconductors Ta, Nb and alloys is set to be 1 since the x, y and z axes are equivalent for these conventional superconductors.

| Material class | Material | $B_{c2}^{\perp}/B_{c2}^{//}$ | $\Delta T/T_c^{onset}$ |
|---|---|---|---|
| | MgTi$_2$O$_4$ | 1.05 | 0.486 |
| **Films** | Pb [32,36,37] | 0.0284 | 0.378 |
| | Ga [33] | 0.0968 | 0.579 |
| | FeSe/SrTiO$_3$ [35,38] | 0.134 | 0.569 |
| **Flakes** | MoS$_2$ [39,40] | 0.0120 | 0.324 |
| | NbSe$_2$ [34] | 0.0165 | 0.328 |
| | ZrNCl [41] | 0.0349 | 0.269 |
| **Bulk** | UPt$_3$ [42,43] | 0.819 | 0.0556 |
| | UTe$_2$ [44] | 0.375 | 0.226 |
| | CeCoIn$_5$ [45-47] | 0.420 | 0.0813 |
| | YBa$_2$Cu$_3$O$_7$ [48,49] | 0.117 | 0.0423 |
| | Nd$_{2-x}$Ce$_x$CuO$_{4-y}$ [50] | 0.0488 | 0.0616 |
| | CsV$_3$Sb$_5$ [51,52] | 0.111 | 0.216 |
| | FeSe [53,54] | 0.604 | 0.165 |
| | MgB$_2$ [55,56] | 0.588 | 0.00993 |
| | LaFeAsO$_{0.89}$F$_{0.11}$ [57,58] | 0.660 | 0.185 |
| | Ta [59] | 1 | 0.000670 |
| | Nb [60] | 1 | 0.0977 |
| | Ag-Sn [61] | 1 | 0.0395 |
| | Au-Pb [61] | 1 | 0.180 |
| | Cu-Sn [61] | 1 | 0.0531 |
| | Tl-Sn [61] | 1 | 0.127 |